\documentclass{aa}
\usepackage{graphics}
\newcommand\spec[3]{\mbox{$^{#1}{#2}_{#3}$}}
\newcommand{\xmm}{{\it XMM-Newton}}

\newcommand\lax{\>\vcenter{\hbox{$<$\hskip-.75em\lower1.0ex\hbox{$\sim$}}}\>}
\newcommand\gax{\>\vcenter{\hbox{$>$\hskip-.75em\lower1.0ex\hbox{$\sim$}}}\>}

\begin{document}

   \title{High resolution X-ray spectroscopy of $\zeta$~Puppis with the 
	XMM-Newton Reflection Grating Spectrometer}
   \author{S. M. Kahn \inst{1}, M. A. Leutenegger \inst{1}, 
          J. Cottam \inst{1}, G. Rauw\inst{2}\thanks{Postdoctoral Researcher
          FNRS (Belgium)}, J.-M. Vreux\inst{2}, A. J. F. den
          Boggende \inst{3}, R.~ Mewe \inst{3},
          \and
          M. G\"{u}del\inst{4}
          }

   \offprints{skahn@astro.columbia.edu}

   \institute{Department of Physics and Columbia Astrophysics Laboratory,
              Columbia University,
              550 West 120th Street, New York, NY 10027, USA
	 \and Institut d'Astrophysique et de G\'{e}ophysique de
              l'Universit\'{e} de Li\`{e}ge, 5, Avenue de Cointe,
              B-4000 Li\`{e}ge, Belgium  
         \and Space Research Organization of the Netherlands, Sorbonnelaan 2,
              3548 CA, Utrecht, The Netherlands 
         \and Laboratory for Astrophysics, Paul Scherrer Institute,
              W\"urenlingen and Villigen, 5232 Villigen PSI, Switzerland 
             }

   \authorrunning{S.\,M.\,Kahn et al.}
   \titlerunning{X-ray spectroscopy of $\zeta$~Puppis.}

   \abstract{
We present the first high resolution X-ray spectrum of the bright O4Ief
supergiant star $\zeta$~Puppis, obtained with the Reflection Grating
Spectrometer on-board \xmm.  The spectrum exhibits bright emission lines of
hydrogen-like and helium-like ions of nitrogen, oxygen, neon, magnesium, and
silicon, as well as neon-like ions of iron.  The lines are all significantly
resolved, with characteristic velocity widths of order $1000\, -\, 1500\, {\rm
km\, s^{-1}}$. The nitrogen lines are especially strong, and indicate that the
shocked gas in the wind is mixed with CNO-burned material, as has been
previously inferred for the atmosphere of this star from ultraviolet spectra.
We find that the forbidden to intercombination line ratios within the
helium-like triplets are anomalously low for N VI, O VII, and Ne IX.  While
this is sometimes indicative of high electron density, we show that in this
case, it is instead caused by the intense ultraviolet radiation field of the
star.  We use this interpretation to derive constraints on the location of the
X-ray emitting shocks within the wind that are consistent with current
theoretical models for this system.  
   }
   \maketitle

\section{Introduction}

The initial discovery of X-ray emission from O stars with the {\it Einstein\
 Observatory} in the late 1970s (Harnden et al. \cite{HarnEL}), sparked a
 vigorous field of research aimed at better understanding the production of
 hot gas in such systems.  Most current models invoke hydrodynamic shocks
 resulting from intrinsic instabilities in the massive, radiatively driven
 winds from these stars (for a detailed review see Feldmeier et
 al. \cite{Feld}).  Support for this picture comes from the fact that the
 X-ray flux is not highly absorbed at low energies (Cassinelli \& Swank
 \cite{CS}; Corcoran et al. \cite{CorEL}), which is expected if the X-ray
 emitting gas is distributed throughout the wind rather than in some form of
 hot corona in the outer atmosphere of the star.  However, the X-ray
 observations to date have not been especially constraining for stellar wind
 models.  This is primarily due to the low spectral resolution of the
 available nondispersive detectors, which has precluded the study of
 individual atomic features so crucial to the unambiguous determination of
 physical conditions in the shocked gas. 

In this Letter, we present one of the first high resolution X-ray spectra of
an early-type star, the O4Ief supergiant $\zeta$~Puppis, which we have
obtained with the Reflection Grating Spectrometer (RGS) experiment on the
\xmm\ Observatory.  $\zeta$~Pup is an excellent target for such work since it
is the brightest O star in the sky, and has consequently been very well
studied at longer wavelengths.  In particular, Pauldrach et
al. (\cite{PaulEL}) have developed a very detailed NLTE model of the
atmosphere and wind of this star, constrained by high quality ultraviolet
spectra from Copernicus and IUE.  They find a mass loss rate of $\sim 5.1
\times 10^{-6}\, {\rm M_{\odot}\, yr^{-1}}$ and an effective temperature
$\sim 42\, 000\, {\rm K}$.  The derived abundances indicate that the
atmosphere is mixed with CNO-processed material, consistent with the
theoretical picture of a highly evolved star near the end of core hydrogen
burning. Hillier et al. (\cite{HilEL}) found that the Pauldrach et al. model
is consistent with the ROSAT X-ray spectrum of $\zeta$~Pup, but only if the
X-ray emission arises in shocks distributed throughout the wind.  At the very
lowest energies ($\leq 200\, {\rm eV}$), they find that the emitting material
must be $\geq 100 \rm R_*$ away from the star.  This is a consequence of the
fact that in the Pauldrach et al. model, helium is mostly only singly ionized
in the outer regions of the wind, so that the photoelectric opacity is very
high at soft X-ray energies.  

Our \xmm\ RGS spectrum is dominated by broad emission lines of mostly
hydrogen-like and helium-like charge states of nitrogen, oxygen, neon,
magnesium, and silicon, and neon-like ions of iron.  The data provide a number
of important constraints on the nature and location of the X-ray emitting
material in the $\zeta$~Pup wind.  Of particular interest is the fact that we
see a suppression of the forbidden line and enhancement of the
intercombination line in the helium-like triplets of nitrogen, oxygen, and
neon.  We show that this is a natural consequence of the intense ultraviolet
radiation field in the wind, and that it allows us to place constraints on the
location of the X-ray emitting shocks relative to the star.  Additional
constraints come from the emission line velocity profiles.  We show that the
data are consistent with the predictions of the Pauldrach et al. model.

In section 2, we describe the details of the RGS observation, the nature of
our data reduction and analysis, and the key observational features of the
spectrum.  In section 3, we consider the implications of our results for our
understanding of $\zeta$~Pup, and of the X-ray emission from O stars in
general.  
  
\section{Observations and Data Analysis}

\begin{figure*}[t]
      \resizebox{17cm}{!}{\rotatebox{90}{\includegraphics{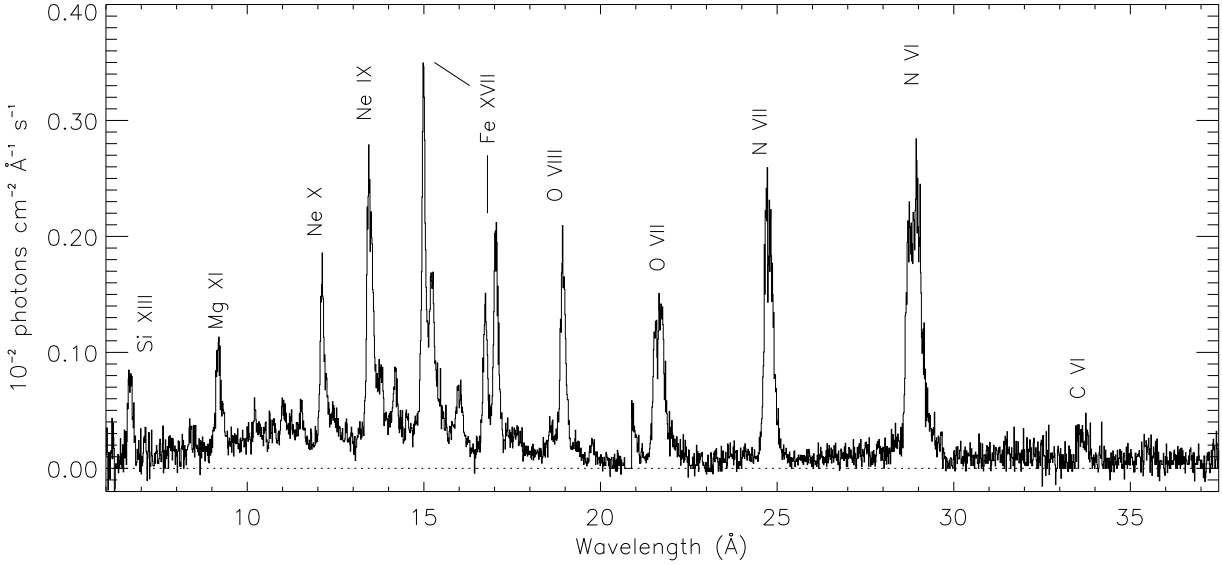}}}
      \caption{First order background subtracted spectrum. It has been corrected for the effective area of the instrument.   
              }
         \label{FigSpectrum}
   \end{figure*}

The RGS covers the wavelength range of $5$ to $35\, {\rm \AA}$ with a
resolution of $0.05\, {\rm \AA}$, and a peak effective area of about $140\,
{\rm cm^2}$ at $15\, {\rm \AA}$. $\zeta$~Pup was observed for 57.4 ks on 2000
June 8. The data were processed with the \xmm~Science Analysis Software
(SAS). Filters were applied in dispersion channel versus CCD pulse height
space to separate the spectral orders, and the source region was separated
with a $1\arcmin$ spatial filter. The background spectrum was obtained by
taking events from a region spatially offset from the source. The wavelengths
assigned to the dispersion channels are based on the pointing and geometry of
the telescope and are accurate to $\sim 0.008\, {\rm \AA}$ (see den Herder et
al.~\cite{denHEL}; this vol.). The effective area was simulated with a
Monte Carlo technique using the response matrix of the instrument and the
exposure maps produced by the SAS. Based on ground calibration, we expect the
uncertainty in the effective area to be less than $10\%$ above $9\, {\rm \AA}$
and at most $20\%$ for shorter wavelengths (den Herder et
al.~\cite{denHEL}). A fluxed spectrum, corrected for effective area, with the
two first order spectra added together to maximize statistics, is presented in
Fig.~\ref{FigSpectrum}. There is a small discontinuity in the spectrum near
the nitrogen Lyman $\beta$ line at $20.91\, {\rm \AA}$ caused by a gap between
two CCD chips. The spectrum was also extracted for analysis in XSPEC using
standard SAS routines.  

Due to a large solar flare event that occurred close to the time of the
$\zeta$~Pup observation, all three EPIC detectors and the Optical Monitor on
\xmm\ were switched off while this source was observed.  Thus, only data from
RGS are available for analysis. 

\subsection{Light Curve}

\begin{figure}
      \resizebox{8.5cm}{!}{\rotatebox{90}{\includegraphics{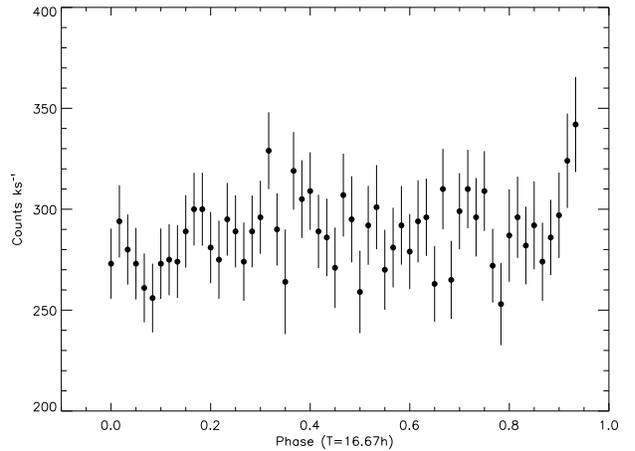}}}
      \caption[]{Light curve of first order events in the range 6.2 to 13.8
      {\AA}, corresponding to 0.9 to 2.0 keV. Events from both instruments
      have been binned together. The time dependence is plotted as a function
      of the 16.67 hr period reported in Bergh\"{o}fer et
      al. 
              }
         \label{FigLightcurve}
   \end{figure}

Previous ROSAT obervations of $\zeta$ Pup (Bergh\"{o}fer et
al.~\cite{Berghofer}) show evidence for a 16.67 h period of variability in
the X-ray emission between 0.9 and 2.0 keV. We have extracted a light curve
(Fig.~\ref{FigLightcurve}) of the events in this energy range in the RGS data set. There is no significant observed variation on this time scale apparent to the
eye, and the fit to a constant intensity yields $\chi^2_{\nu}=0.9$. However,
our upper limit to the percentage variation at this period is not inconsistent
with the Bergh\"{o}fer et al. detection.


\subsection{Emission Line Intensities and Emission Measure Analysis}

The spectrum displayed in Fig.~\ref{FigSpectrum} is composed almost entirely of emission lines, with a weak underlying continuum. The lines are all broadened,
with characteristic velocity widths of order $1000\, {\rm km\, s^{-1}}$. In nitrogen,
oxygen, and neon the He-like forbidden lines are very weak and the
intercombination lines are bright. The line fluxes (Table~\ref{TabFlux}) were
measured by taking the integrated flux of the spectrum over the line and
subtracting a corresponding amount of continuum flux. Lines that originate
from the same ion and that are blended were evaluated as one complex (for
example, the He-like triplets or the Fe XVII emission around 15 or 17
{\AA}). The continuum strength was determined by taking the flux of a spectral
region free of lines but near the line in question. In cases where this was
not possible, the continuum strength was interpolated from the strength in
other regions of the spectrum. Some of the measurements were complicated by
the presence of overlapping lines. When possible, the flux of these other
lines was estimated from the flux of lines originating from the same ion by
comparing with line power ratios.

\begin{figure}
      \resizebox{8.5cm}{!}{\rotatebox{90}{\includegraphics{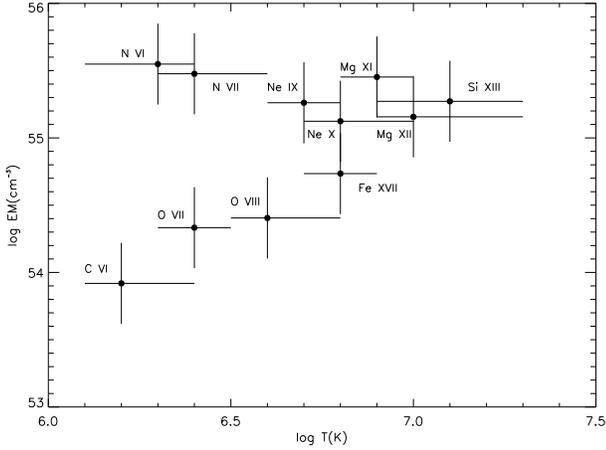}}}
      \caption{Emission measure calculated from each line as a function of
      ${T_{\rm form}}$. The horizontal error bars indicate the temperature
      range over which $P_{\rm line}f_{\rm i}$ is at least half its maximum
      value. Solar abundances are assumed.  
              }
         \label{FigEM}
   \end{figure}

We calculated the emission measure (see Fig.~\ref{FigEM}) for each ion
assuming solar abundances (Anders \& Grevesse \cite{Anders}) and a temperature
given by the temperature of formation for the dominant lines from that charge
state. The emission measure is  
\begin{equation}
EM = F_{\rm line}\frac{4\pi d^2}{P_{\rm line}A f_{\rm i}} 
\end{equation}
where $F_{\rm line}$ is the observed flux in the line, $d$ is the distance to
the source, $P_{\rm line}$ is the line power, $A$ is the elemental abundance,
and $f_{\rm i}$ is the ion fraction evaluated at the temperature of formation.
We used line powers from the APEC code (Smith \& Brickhouse \cite{Smith}),
which includes ion fractions from Mazzotta et al. (\cite{Mazzotta}). We take
$d=450\,{\rm pc}$ (Schaerer et al.~\cite{Sch}). The line fluxes must be 
corrected for interstellar absorption, although this is a minor effect. We
take $N_{\rm H}=10^{20}\,{\rm cm^{-2}}$ (Chlebowski et al.~\cite{Chle}) and we
use cross sections from Morrison \& McCammon~(\cite{Morrison}). 

We find that the emission measures derived from the nitrogen emission lines
are at least an order of magnitude greater than those for carbon and oxygen,
which both have temperature of formation ranges that overlap with that of
nitrogen. This indicates that the ratios ${A_{\rm N}/A_{\rm O}}$ and ${A_{\rm
N}/A_{\rm C}}$ are substantially higher than solar, even allowing for a factor
of two uncertainty due to the crudeness of the emission measure analysis. This
result is consistent with the inference of atmospheric abundances by Pauldrach
et al. (\cite{PaulEL}), based on their analysis of UV spectra.  It indicates
that the wind material in $\zeta$~Pup has been significantly mixed with matter
that has undergone CNO burning in the stellar interior. 
  
The emission measure expected for a smooth, spherically symmetric wind with
parameters appropriate to $\zeta$~Pup is $EM=6.5\times 10^{60} {\rm cm^{-3}}$.
The fact that we find EMs that are lower by 4 to 5 orders of magnitude implies
that only a small fraction of the wind material is heated to X-ray emitting
temperatures.        
     
\begin{table}
\begin{center}
\caption{Measured fluxes for prominent emission lines with rest wavelengths in
{\AA}. \label{tbl-1}}
\begin{tabular}{cc}
\hline
\hline
Line & Flux \\
     & (${\rm photons\, cm^{-2}\, s^{-1}}$) \\
\hline
Si XIII ($6.65,6.69,6.74$) & $(1.1\pm0.4)\times 10^{-4}$ \\
Mg XII ($8.42$) & $(4.1\pm2.8)\times 10^{-5}$ \\
Mg XI  ($9.17,9.23,9.31$) & $(2.0\pm0.7)\times 10^{-4}$ \\
Ne X   ($12.13$) & $(1.8\pm0.7)\times 10^{-4}$ \\
Ne IX  ($13.45,13.55,13.70$) & $(5.0\pm1.0)\times 10^{-4}$ \\
Fe XVII ($15.01,15.26$) & $(7.4\pm0.7)\times 10^{-4}$ \\
Fe XVII ($16.78,17.05,17.10$) & $(4.7\pm0.8)\times 10^{-4}$ \\
O VIII ($18.97$) & $(3.50\pm0.22)\times 10^{-4}$ \\
O VII  ($21.60,21.80,22.10$) & $(5.1\pm0.4)\times 10^{-4}$ \\
N VII  ($24.78$) & $(6.4\pm0.5)\times 10^{-4}$ \\
N VI ($28.78,29.08,29.53$) & $(1.13\pm0.11)\times 10^{-3}$ \\
C VI   ($33.74$) & $(6.4\pm2.8)\times 10^{-5}$ \\
\hline
\label{TabFlux}
\end{tabular}

\end{center}
\end{table}


\subsection{Continuum emission analysis}

The continuum emission has a very low intensity relative to the bright line
emission. To ensure that the continuum that we see in the spectrum is real,
and not, for example, an artifact of faulty background subtraction, we plotted
the first order events in a region free of emission lines as a function of
their spatial distribution in the cross-dispersion direction, without
subtracting the background. The peak at the location of the source spectrum is
clearly visible, and indicates the presence of true continuum, well above
background.  The detected continuum also significantly exceeds the scattered
light contribution from the ensemble of detected emission lines.  

We looked for evidence of discrete photoelectric absorption edges in the
continuum. Although there are no edges evident, the constraints are weak due
to the low strength of the continuum emission. The upper limits on the optical
depths are given in Table ~\ref{TabEdges}.  These upper limits are
incompatible with the detection of a strong edge feature near 0.6 keV reported
by Corcoran et al. (\cite{CorEL}), but that is perhaps not surprising,
considering the complexity of the spectrum and the low spectral resolution of
their measurements. 

The continuum is so weak that it is difficult to ascertain its overall
shape, especially in the region from 8 - 17~{\AA} where there are many lines.
We tried fitting with a bremsstrahlung model in XSPEC, but the results were
inconclusive.  


\begin{table}
\begin{center}
\caption{Upper limits on the strengths of the K-edges of Ne, O, and N. \label{tbl-2}}
\begin{tabular}{ccc}
\hline
\hline
Edge & Wavelength & Upper limit on $\tau_0$ \\
     & ({\AA}) & \\
\hline
Ne X    & $9.10$ & $0.5$ \\
Ne IX   & $10.37$ & $0.5$ \\
O VIII  & $14.23$ & $0.2$ \\
O VII   & $16.78$ & $0.3$ \\
N VII   & $18.59$ & $0.5$ \\
N VI    & $22.46$ & $0.5$ \\
\hline
\label{TabEdges}
\end{tabular}
\end{center}
\end{table}


\subsection{He-like triplet ratios}
\begin{figure}
      \resizebox{8.5cm}{!}{\rotatebox{270}{\includegraphics{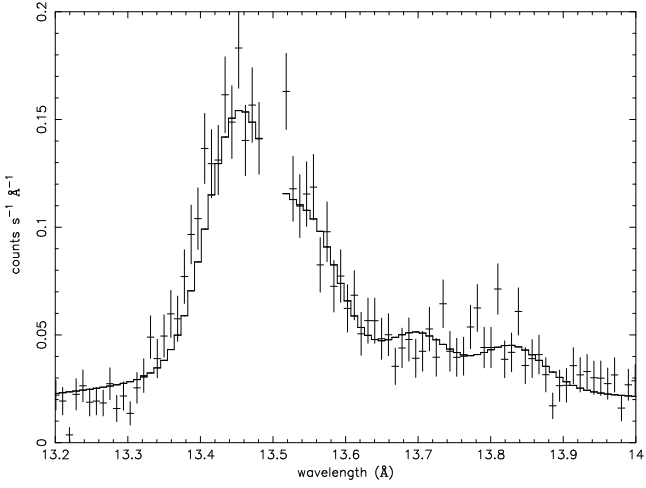}}}
      \caption[]{The Ne IX triplet with best fit model. Only RGS 2 data are
      shown for clarity, but both were used in the fit. There is also a Fe
      XVII line at $13.825\, {\rm \AA}$. The rest wavelengths of the three
      lines are at 13.45 {\AA} (r), 13.55 {\AA} (i), and 13.70 {\AA} (f).
              }
         \label{FigNeIX}
   \end{figure}

The forbidden to intercombination line ratios, ${R = f/i}$, for helium-like
oxygen and neon were obtained by fitting in XSPEC. In each case, we fit a
gaussian to the Ly$\alpha$ line of the respective element, and used that
profile for each of the three components in the He-like triplet. We also take
into account both the low level continuum and lower flux lines in the near
vicinity. For Ne IX we find R = $0.34\pm0.11$ and for O VII we find R =
$0.19\pm0.08$. The fit to Ne IX is shown in Fig.~\ref{FigNeIX}. These values
are well below the expected values for low density plasmas in collisional
equilibrium (see Mewe et al. \cite{MeweEL}).  A similar analysis did not work
very well for the N VI He-like triplet because of the complexity of both the
Ly$\alpha$ and He-like profiles for that element.  Nevertheless, it is clear
from the data that the forbidden line is strongly suppressed for N VI as well.
The Si XIII and Mg XI triplets are too blended to allow the forbidden line and
intercombination line to be quantitatively separated. Again, it is clear from
the data that the Mg XI forbidden line is suppressed, although not as strongly
as for Ne IX. 

The conversion of forbidden line to intercombination line emission in high
density plasmas is a well known effect. However, this can also occur at much
lower densities if the plasma is exposed to a strong UV radiation field (Mewe
\& Schrijver \cite{Mewe}). To produce a low $R$ ratio, electrons populating
the 2~\spec{3}{S}{1} state must be excited into the 2~\spec{3}{P}{} state. The
state 2~\spec{3}{S}{1} is metastable, but it is easily populated at
collisional equilibrium temperatures, and the corresponding emission line
intensity is normally strong. When the excitation rate to the 2~\spec{3}{P}{}
levels becomes comparable to the decay rate to the ground state, the forbidden
line is suppressed. In a high density plasma, this occurs because the
collisional excitation rate from 2~\spec{3}{S}{1} to 2~\spec{3}{P}{} competes
effectively with radiative decay. However, since this is a dipole transition,
it can also be photoexcited if the ambient UV flux at the appropriate
wavelength is sufficiently high. To calculate the photoexcitation rate, we
estimated the emission from $\zeta$~Pup for the frequencies of the
2~\spec{3}{S}{1} to 2~\spec{3}{P}{} transitions for N, O, Ne, Mg, and Si. We
assumed a blackbody spectrum with ${T_{\rm eff}}\, =\, 42\, 000 {\rm K}$
(Pauldrach et al. \cite{PaulEL}). The rate of photoionization is given by
${R_{\rm PE}=F_{\rm \nu}\frac{\pi e^2}{mc}f}$, where $f$ is the oscillator
strength. We used oscillator strengths from Cann and Thakkar (\cite{Cann}) and
Sanders and Knight (\cite{Sanders}). The flux is given by $F_{\rm
\nu} = 2\pi (1-\sqrt{1-(\frac{R_*}{R})^2})I_{\rm \nu}$, where $I_{\rm \nu}$ is
the specific intensity for the blackbody. 

We used the archived IUE UV spectrum of $\zeta$~Pup in addition to the
Copernicus UV spectrum (Morton and Underhill~\cite{MU}) to assess the validity
of our blackbody model for the UV emission from the photosphere. We compared
the measured flux at the wavelengths of the 2~\spec{3}{S}{1} to
2~\spec{3}{P}{} transitions for N VI, O VII, and Ne IX to the flux predicted
by the blackbody model. After correcting the measured flux for absorption,
these agree to within a factor of two. Since the critical radius where
${R_{\rm PE}}={R_{\rm decay}}$ depends approximately on $\sqrt{F_{\rm \nu}}$,
the radii we calculate are valid to within a factor of $\sqrt{2}$. The
measured UV spectra also indicate that there is negligible optical depth in
these lines due to the wind. This is expected since these are excited state
transitions, and since the helium-like plasma only represents a very small
fraction of the wind.

In Table~\ref{TabPhotoexcitation} we list the decay rates of  2~\spec{3}{S}{1}
and the photoexcitation rates from 2~\spec{3}{S}{1} to 2~\spec{3}{P}{}
evaluated at the photosphere. It is clear from this calculation that the
forbidden line suppression we observe is in fact due to photoexcitation from
$\zeta$ Pup's high UV flux, and not due to collisional excitation at high
densities, as long as the emitting regions are close enough that the UV is not
sufficiently diluted. We can thus place constraints on the location of shock
formation for each of the observed lines. Since the forbidden lines in N, O,
and Ne are strongly suppressed, the emission must occur at radii smaller
than the radii at which the photoexcitation and decay rates are equal (the
critical radius). Since Mg XI has a somewhat suppressed forbidden
line, its emission probably occurs near the critical radius. The emission from
Si must occur farther out than the critical radius.    
 
\begin{table}
\begin{center}
\caption{Comparison of photoexcitation and decay rates of the 2~\spec{3}{S}{}
        state. Decay rates are from Drake (\cite{Drake}). \label{tbl-3}
        }
\begin{tabular}{cccc}
\hline
\hline
Ion & Decay Rate  & Photoexcitation & Radius \\
 & & Rate ($R=R_*$) & at which\\
 & 2~\spec{3}{S}{} $\rightarrow$ 1~\spec{1}{S}{} & 
2~\spec{3}{S}{} $\rightarrow$ 2~\spec{3}{P}{} & ${R_{\rm PE}=R_{\rm decay}}$\\
    & (${\rm s^{-1}}$) & (${\rm s^{-1}}$) & ($R_*$)\\
\hline
Si XIII & $3.56\times 10^5$ & $4.83\times 10^6$ & $2.7$ \\
Mg XI   & $7.24\times 10^4$ & $7.36\times 10^6$ & $7.1$ \\
Ne IX   & $1.09\times 10^4$ & $1.11\times 10^7$ & $22.6$ \\
O VII   & $1.04\times 10^3$ & $1.66\times 10^7$ & $89$ \\
N VI    & $2.53\times 10^2$ & $1.99\times 10^7$ & $198$ \\
\hline
\label{TabPhotoexcitation}
\end{tabular}
\end{center}
\end{table}


\subsection{Line profile analysis}
\begin{figure}
      \resizebox{8.5cm}{!}{\rotatebox{90}{\includegraphics{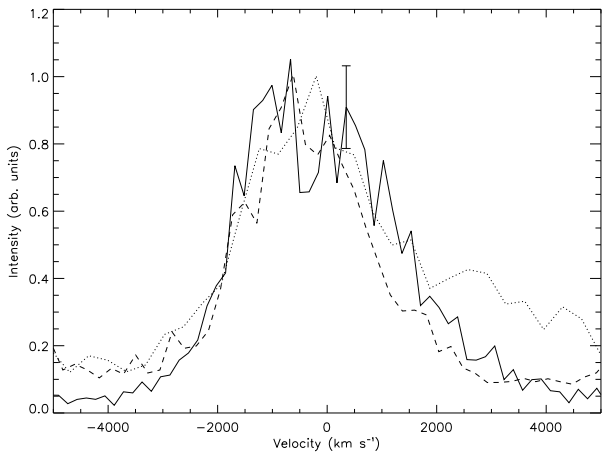}}}
      \caption[]{Ly $\alpha$ lines in velocity space. Nitrogen is the solid
      line, oxygen is dashed, and neon is dotted. The intensities have been
      renormalized for comparative purposes. The error bar is representative
      for the nitrogen line.  Note the discrete structure in the nitrogen
      peak. The shoulder on the Ne X line is not a velocity feature; it is an
      emission line from Fe XVII.     
              }
         \label{FigVelocity}
   \end{figure}

The projected velocity ($v_{\rm p}$) profile for a thin spherical shell (with
a single radial velocity) is flat, or $\frac{dI}{dv_{\rm p}}=({\rm
const.})$. We expect the emission line profile to appear as a convolution of
the radial emission intensity with this flat projected velocity profile. The
velocity of each shell is given as a function of radius by the conventional 
$\beta$-model:  ${v(r) \approx v_{\rm \infty} (1-r_{\rm 0}/r)^{\beta}}$
(Lamers \& Cassinelli \cite{Lamers}) where $\beta \approx 0.8$ and ${r_{\rm 0}
\approx R_{\rm *}}$. Lines formed at larger radii will therefore appear
broader than lines formed close to the star. Furthermore, lines originating
from larger radii are formed over a region with a small velocity
gradient. Therfore, we expect these lines to appear more flat-topped than
lines originating closer to the star. This is apparent in the observed
Ly$\alpha$ lines, which are plotted in velocity space in
Fig.~\ref{FigVelocity}. The N VII peak is noticeably broader and has a
substantially different shape than the other peaks in the plot. It also shows
evidence for resolved, discrete structure, given the resolution of the
instrument. Note that other lines overlap with the Ne X line; it does not have
a red shoulder.  

In Table~\ref{TabVelocity} we list the shifts in the line centroids and the
line widths in velocity space for Ne X, O VIII, and N VII. The fit to N VII is
poor due to the complex structure in the line profile.


\begin{table}
\begin{center}
\caption{Velocity widths and shifts of the Lyman $\alpha$ lines. \label{tbl-4}
        }
\begin{tabular}{ccc}
\hline
\hline
Ion & Velocity shift & Velocity width \\
    & (${\rm km\, s^{-1}}$) & (${\rm km\, s^{-1}}$)\\
\hline
Ne IX   & $250\pm125$ & $940\pm150$ \\
O VII   & $400\pm80$ & $1230\pm80$ \\
N VI    & $0\pm60$ & $1370\pm100$ \\
\hline
\label{TabVelocity}
\end{tabular}
\end{center}
\end{table}

\section{Discussion}

The RGS spectrum we have presented provides very strong confirmation for a
number of key aspects of conventional models of X-ray emission from early-type
stars in general, and for $\zeta$~Pup in particular.  The emission is
dominated by broad emission lines indicative of hot plasma that is flowing
outward with the wind.  We do not see evidence for strong attenuation at low
energies, which confirms that the X-ray emitting regions are not concentrated
at the base of the wind, but instead are distributed out to large radii.
These very simple results agree well with the predictions of wind instability
models. 

As we have shown, the suppression of the forbidden lines in the He-like
triplets of low Z elements allows us to derive upper limits to the radii of
the emitting shocks in each case.  It is interesting to compare these values
with Fig. 1 of Hillier et al. (\cite{HilEL}), which is a plot of the radius at
which optical depth unity is reached in the wind as a function of X-ray
energy for the NLTE model of $\zeta$~Pup calculated by Pauldrach et
al. (\cite{PaulEL}).  At the energies of the N VI, O VII, Ne IX, Mg XI, and Si
XIII lines, unit optical depth is achieved at 22, 23, 9, 3.5, and 2.5 stellar
radii, respectively, in this model.  These values are quite compatible with
both our derived upper limits for N VI, O VII, Ne IX, and Mg XI, and our
derived lower limit for Si XIII. Since the density in the wind drops off like
$r^{-2}$, and the emissivity is proportional to $n^{2}$, we expect radii
characteristic of the smallest radius at which 
the overlying wind is still transparent to the respective line, in each case.
Our results are consistent with this expectation.  

Further support for this picture comes from the observed velocity profiles.
The higher Z lines have characteristic widths $\sim \, 1000\, {\rm km\,
s^{-1}}$, whereas the N VII line is distinctly broader.  The terminal velocity
in the $\zeta$~Pup wind is $2260\, {\rm km\, s^{-1}}$ (Groenewegen et
al. \cite{GroeEL}), so the higher Z lines are most likely emitted at only a
few stellar radii, whereas N VII can come from considerably further out.   

We have shown that the respective line intensities are suggestive of
significant enhancements of the nitrogen abundance relative to carbon and
oxygen, as one would expect for CNO-processed material.  This, again, agrees
well with the Pauldrach et al. model. Meynet \& Maeder (\cite{MeMa}) recently
presented new evolutionary models for rotating single stars. They found that
rotational mixing produces a significant surface helium and nitrogen
enhancement. Meynet \& Maeder suggested that stars with enhanced
He-abundances and large projected rotational velocities are natural
descendants of very fast rotating main-sequence stars. In these stars, the
chemical enrichment at the surface is very fast and as a result of the strong
rotational mixing the chemical structure of these stars could be near
homogeneity. With its large projected rotational velocity of $v\,\sin{i}
\simeq 203$\,km\,s$^{-1}$ (Penny \cite{Pen}), $\zeta$\,Pup most probably falls
into this category. 

\begin{acknowledgements}
  We acknowledge useful comments from the referee, M. Corcoran, which
  significantly improved the presentation of this paper.
  This work is based on observations obtained with \xmm, an ESA science
  mission with instruments and contributions directly funded by ESA Member
  States and the USA (NASA).  The Columbia University team is supported by
  NASA.  SRON is supported by the Netherlands Foundation for Scientific
  Research (NWO). GR and JMV acknowledge support by the SSTC-Belgium under
  contract P4/05 and by the PRODEX XMM-OM Project. JC acknowledges the support
  of a NASA Graduate Student Researchers Program Fellowship. 
\end{acknowledgements}




\begin{thebibliography}{}

   \bibitem[1989]{Anders} Anders E., Grevesse N., 1989, 
      Geochimica et Cosmochimica Acta 53, 197 

   \bibitem[1996]{Berghofer}Bergh\"{o}fer T.W., Baade D., Schmitt J.H.M.M.,
      Kudritzki R.-P., Puls J., Hillier D.J., Pauldrach A.W.A., 1996, A\&A
      306, 899

   \bibitem[1992]{Cann} Cann N.M., Thakkar A.J., 1992, Phys. Rev. A 46, 9,
      5397

   \bibitem[1983]{CS} Cassinelli J.P., Swank J.H., 1983, ApJ 271, 681

   \bibitem[1989]{Chle}Chlebowski T., Harnden F. R., Sciortino S., 1989, ApJ 341, 427
   \bibitem[1993]{CorEL} Corcoran M. F., Swank J. H., Serlemitsos P. J., Boldt
   E., Petre R., Marshall F. E., Jahoda K., Mushotzky R., Szymkowiak A.,
   Arnaud K.,  Smale A. P., Weaver K., Holt S. S., 1993, ApJ 412, 792

   \bibitem[2000]{denHEL} den Herder, J.W., et al., 2000, this volume.

   \bibitem[1971]{Drake} Drake G.W.F., 1971, Phys. Rev. A 3, 908

   \bibitem[1997]{Feld} Feldmeier A., Kudritzki R.-P., Palsa R., Pauldrach
   A.W.A., Puls J., 1997, A\&A 320, 899

   \bibitem[1989]{GroeEL} Groenewegen M.A.T., Lamers H.J.G.L.M., 1989, A\&AS
   79, 359

   \bibitem[1979]{HarnEL} Harnden F.R. Jr., Branduardi G., Elvis M., et al.,
   1979, ApJ 234, L51-54

   \bibitem[1993]{HilEL} Hillier D.J., Kudritzki R.-P., Pauldrach A.W.A.,
   Baade D., Cassinelli J.P., Puls J., Schmitt J.H.M.M., 1993, A\&A 276,
   117 

   \bibitem[1999]{Lamers} Lamers H.J.G.L.M., Cassinelli J.P., 1999,
      Introduction to Stellar Winds 

   \bibitem[1998]{Mazzotta}Mazzotta P., Mazzitelli G., Colafrancesco S.,
   Vittorio N., 1998, A\&AS 133, 403

   \bibitem[1978]{Mewe}Mewe R., Schrijver J., 1978, A\&A 65, 99

   \bibitem[2000]{MeweEL} Mewe R., Porquet D., Raassen A.J.J., Kaastra J.S.,
   Dubau J., 2000, in preparation

   \bibitem[2000]{MeMa} Meynet G., Maeder A., 2000, A\&A 361, 101

   \bibitem[1977]{MU} Morton D.C., Underhill A.B., ApJS 33, 83

   \bibitem[1983]{Morrison} Morrison R., McCammon D., 1983, 
      ApJ 270, 119 

   \bibitem[1994]{PaulEL}Pauldrach A.W.A., Kudritzki R.P., Puls J., Butler
   K., Hunsinger J., 1994, A\&A 283, 525

   \bibitem[1996]{Pen} Penny L.R., 1996, ApJ 463, 737
 
   \bibitem[1989]{Sanders}Sanders F. C., Knight R. E., 1989, Phys. Rev. A 39,
      9, 4387

   \bibitem[1997]{Sch}Schaerer D., Schmutz W., Grenon M., 1997, ApJ 484, L153

   \bibitem[2000]{Smith}Smith R., Brickhouse N., 2000, Astrophysical Plasmas:
   Codes, Models, and Observations, Proceedings of the conference held in
   Mexico City, October 25-29, 1999, Eds. Jane Arthur, Nancy Brickhouse, and
   Jos\'{e} Franco, Revista Mexicana de Astronom\'{i}a y Astrof\'{i}sica
   (Serie de Conferencias), Volume 9, p. 134-136

\end{thebibliography}
\end{document}